\documentclass[11pt,a4paper]{article}
\usepackage[T1]{fontenc}
\usepackage[utf8]{inputenc}
\usepackage{lmodern}
\usepackage[margin=1in]{geometry}
\usepackage{microtype}
\usepackage{amsmath,amssymb,amsthm,mathtools}
\usepackage{booktabs,tabularx}
\usepackage{algorithm}
\usepackage{algpseudocode}
\usepackage[numbers,square]{natbib}
\usepackage{xurl}
\usepackage[hidelinks]{hyperref}
\usepackage[capitalise,nameinlink,noabbrev]{cleveref}
\usepackage{fancyhdr}
\usepackage{needspace}
\usepackage{aliascnt}

\hypersetup{
  pdftitle={Online Metric Matching with Unknown i.i.d. Arrivals},
  pdfsubject={Sample-free stochastic online metric matching in general metric spaces},
  pdfkeywords={online metric matching, unknown iid, optimal transport, Fair-Bias}
}
\numberwithin{equation}{section}
\newtheorem{theorem}{Theorem}[section]
\newaliascnt{lemma}{theorem}
\newtheorem{lemma}[lemma]{Lemma}
\aliascntresetthe{lemma}
\newaliascnt{proposition}{theorem}
\newtheorem{proposition}[proposition]{Proposition}
\aliascntresetthe{proposition}
\newaliascnt{corollary}{theorem}
\newtheorem{corollary}[corollary]{Corollary}
\aliascntresetthe{corollary}
\theoremstyle{remark}
\newaliascnt{remark}{theorem}

\aliascntresetthe{remark}
\crefname{theorem}{Theorem}{Theorems}
\crefname{lemma}{Lemma}{Lemmas}
\crefname{proposition}{Proposition}{Propositions}
\crefname{corollary}{Corollary}{Corollaries}
\crefname{remark}{Remark}{Remarks}
\crefname{appendix}{Appendix}{Appendices}
\newcommand{\E}{\mathbb E}
\newcommand{\Prob}{\mathbb P}
\newcommand{\OPT}{\mathrm{OPT}}
\newcommand{\ALG}{\mathrm{ALG}}
\newcommand{\FB}{\textnormal{\textsc{Fair-Bias}}}

\newcommand{\ind}{\mathbf 1}

\newcommand{\cU}{\mathcal U}

\newcommand{\paperauthors}{%
  Tsubasa Harada\\
  Institute of Science Tokyo\\
  \texttt{harada.t.30af@m.isct.ac.jp}
}
\title{\textbf{Online Metric Matching with Unknown i.i.d. Arrivals}}
\author{\paperauthors}
\date{\today}

\begin{document}
\maketitle
\begin{abstract}
We study online metric matching with $n$ servers
and $n$ requests drawn independently from an unknown distribution.
Each request must be matched immediately and irrevocably to an
unused server, and the algorithm receives no samples before the
process starts.
We give a randomized algorithm that is
$O((\log\log\log n)^2)$-competitive on arbitrary metric spaces
and $O(1)$-competitive on tree metrics.
These guarantees match the corresponding
bounds for known i.i.d.\ request distributions in \cite{GGPW19} up to constant factors.
More generally, our competitive ratio is $O(\alpha(n))$ whenever
the \FB\ algorithm introduced in that work has competitive ratio $\alpha(n)$
for known i.i.d.\ arrivals on the same class of metrics.
Consequently, proving that \FB\ is constant-competitive
on general metrics would also yield a constant-competitive
algorithm for unknown i.i.d.\ arrivals.
\end{abstract}

\section{Introduction}\label{sec:intro}
In online metric matching, $n$ servers are available in a metric
space before the process begins, and $n$ requests arrive sequentially. Each
request must be assigned to an unused server as soon as it arrives. Assignments
cannot be changed later, and the objective is to minimize the total
distance between requests and their assigned servers.
The performance of an algorithm is measured by its competitive ratio,
which compares its cost with the minimum cost of a perfect matching
between all requests and all servers. Informally, an algorithm is
$\alpha$-competitive if its cost is at most $\alpha$ times the cost of a
minimum-cost perfect matching.

Under adversarial arrivals, the
optimal deterministic competitive ratio for general metrics is $2n-1$
\cite{KP93,KMV94}, while randomized algorithms achieve $O(\log^2 n)$
\cite{BBGN14}.

When requests arrive at random, there are three natural models to consider.
In the \emph{known-i.i.d. model},
requests are drawn independently from a common distribution that is given to
the algorithm before any requests arrive. In the \emph{unknown-i.i.d. model},
requests are generated in the same way, but the distribution is not given to
the algorithm. In the \emph{random-order model}, an adversary fixes
the request locations in advance without revealing them to the algorithm.
The requests then arrive in a uniformly random order, and each location is
revealed only upon arrival.

The known-i.i.d. model is no harder than the unknown-i.i.d. model, since an
algorithm can ignore the supplied distribution. The unknown-i.i.d. model is
in turn no harder than the random-order model: conditional on the multiset
of i.i.d.\ requests, their order is uniformly random. Thus, any algorithm
that is $\alpha$-competitive in the random-order model is also
$\alpha$-competitive in the unknown-i.i.d. model.

For the known i.i.d.\ model, \cite{GGPW19} obtain an
$O((\log\log\log n)^2)$ competitive ratio for general metrics and $O(1)$ for tree metrics, whereas the optimal ratio in the
random-order model is $\Theta(\log n)$ \cite{Rag16}.
This gap raises a question also posed in \cite[Section~5]{GGPW19}:
in terms of achievable competitive ratios, is the intermediate unknown-i.i.d.\
model closer to the known-i.i.d.\ model or the random-order model?
For line metrics, \cite{VMF25} report an $O(1)$-competitive algorithm
in the unknown-i.i.d.\ model.\footnote{
Our account is based solely on the symposium abstract.
We have not confirmed the full model assumptions or technical details.
}
Their bound matches the known-i.i.d.\ bound on the line, providing a partial
answer to this question.

\subsection{Our result}
In this paper, we give an algorithm for the unknown-i.i.d.\ model that matches the
known-i.i.d.\ upper bound for general metrics.
Let $\ALG$ denote the total matching cost incurred by the online algorithm,
and let $\OPT$ denote the minimum cost of a perfect matching between all
requests and servers.

\begin{theorem}\label{thm:main}
For every metric space and every placement of $n$ labeled servers, there is a
randomized online algorithm that, for every unknown request distribution $D$,
satisfies
\[
  \E[\ALG]\le O((\log\log\log n)^2)\,\E[\OPT]
\]
when the $n$ requests are drawn i.i.d.\ from $D$. Here, the expectation is over both the requests and the algorithm's random choices.
\end{theorem}

More generally, our competitive ratio is $O(\alpha(n))$ whenever
\FB\ has competitive ratio $\alpha(n)$ for known i.i.d.\ arrivals on
the same class of metrics; see \cref{thm:core-reduction} and the reduction
in \cref{subsec:model}. Thus, if the competitive ratio of \FB\ on general metrics is proved to be $O(1)$, our algorithm would also be $O(1)$-competitive.

The resulting bound is $o(\log n)$, so it separates unknown i.i.d.\ arrivals
from random-order arrivals. It matches the order of
the known-i.i.d.\ upper bound.
On tree metrics, our algorithm is constant-competitive; see \cref{cor:trees}.
The guarantees for the three arrival models are summarized in \cref{tab:models}.

Note that our results do not imply that the optimal competitive ratios in
the known-i.i.d.\ and unknown-i.i.d.\ models have the same asymptotic order.
For example, a separation between the two models could still occur on general metrics if an algorithm
other than \FB\ were $O(1)$-competitive in the known-i.i.d.\ model while
\FB\ had competitive ratio $\omega(1)$.

\begin{table}[H]
\centering
\caption{Competitive-ratio bounds for general and tree metrics.}
\label{tab:models}
\begin{tabularx}{\linewidth}{@{}lccX@{}}
\toprule
Arrival model & General metrics & Tree metrics & Source\\
\midrule
Known i.i.d. & $O((\log\log\log n)^2)$ & $O(1)$ & \cite{GGPW19}\\
Unknown i.i.d. & $O((\log\log\log n)^2)$ & $O(1)$ & \cref{thm:main,cor:trees}\\
Random order & $\Theta(\log n)$ & $\Theta(\log n)$ & \cite{Rag16}\\
\bottomrule
\end{tabularx}
\end{table}

\subsection{Algorithm overview}
Let $\delta_x$ denote the point mass at $x$ and write
\[
  \nu=\frac1n\sum_{j=1}^n\delta_{s_j}
\]
for the distribution obtained by choosing a server label uniformly, where $s_j$
is the position of server $j$. For a known request distribution $D$, the
reduction in \cite[Lemma~2.1]{GGPW19} uses an optimal transport
plan from $D$ to $\nu$ to generate virtual requests from $\nu$. This plan can
be fixed before any requests arrive.

Our algorithm replaces $D$ in this construction by the empirical distribution
\[
  \widehat D_t=\frac1t\sum_{i=1}^t\delta_{r_i},
\]
which includes the current request $r_t$. At each time $t$, it computes an
optimal transport plan from $\widehat D_t$ to $\nu$. Using the conditional
probabilities in the row corresponding to the current request, it draws a virtual server label and
passes it to \FB. The actual request $r_t$ is assigned to the server chosen
by \FB.

Although the transport plan changes as requests arrive, we prove that the
generated virtual labels are independent and each is uniformly distributed
over the $n$ server labels; see \cref{lem:uniform-virtual}. Consequently, their
positions form independent samples from $\nu$, so the guarantees of \FB\ apply
to its execution on the virtual requests. It remains to bound the distances
between actual and virtual requests. We show that their expected sum is at most
$O(\alpha(n))\E[\OPT]$, where $\alpha(n)$ is the competitive ratio of \FB\ for
independent requests from $\nu$; see \cref{thm:core-reduction}.

\subsection{Related work}
\paragraph{Adversarial online metric matching.}
\cite{KP93} and \cite{KMV94} independently give deterministic
$(2n-1)$-competitive algorithms and matching lower bounds.
For randomized algorithms, an $\Omega(\log n)$ lower bound is stated in
\cite{KP98}. \cite{MNP06} give an $O(\log^3 n)$-competitive algorithm,
and \cite{BBGN14} improve this bound to $O(\log^2 n)$.

\paragraph{Random order and known i.i.d.\ arrivals.}
\cite{Rag16} gives an $O(\log n)$-competitive algorithm and proves a matching
$\Omega(\log n)$ lower bound in the random-order model.
For known i.i.d.\ arrivals, \cite{GGPW19} introduce
\FB\ and obtain the $O((\log\log\log n)^2)$ general-metric bound, together with
constant bounds on trees. \cite{DSX22} study uniform metric spaces and provide
a more detailed competitive-ratio analysis under random-order arrivals.

\paragraph{In Euclidean spaces.}
\cite{BFP23} study greedy matching on the line.
They prove constant competitiveness when all server and request locations are
sampled independently from the uniform distribution on $[0,1]$. When the server
locations are chosen adversarially and the requests are still sampled
independently from the uniform distribution on $[0,1]$, they obtain an
$O(\log n)$ competitive ratio. \cite{YY26} obtain
$O(1)$-competitive algorithms for online metric matching in the
known-i.i.d.\ model on $[0,1]^d$, under the uniform distribution in any fixed
dimension and under distributions with densities bounded above by a constant
in fixed dimensions $d\ge3$.
\cite{LVY26} analyze online metric matching with independent requests whose
distributions need not be identical.

\paragraph{Separation between arrival models in other matching problems.}
\cite{ABCG26} separate random-order and unknown-i.i.d.\ arrivals for a different
matching problem. They study non-bipartite minimum-cost matching on the line,
where all participants arrive online. Upon each arrival, the algorithm either matches
the new participant to a previously unmatched participant or leaves it
unmatched for a future arrival. They obtain an $O(\log^2 n)$ competitive ratio
under unknown i.i.d.\ arrivals, whereas the competitive ratio is unbounded
under random-order arrivals.

\Needspace{6\baselineskip}
\section{Preliminaries}\label{sec:prelim}
\subsection{The model and benchmark}\label{subsec:model}
Let $(X,d)$ be an arbitrary metric space and let $S=(s_1,\ldots,s_n)$ be the
known server placement, with labels in $[n]=\{1,\ldots,n\}$. Distinct server
labels may occupy the same position. Each server can be used once. The requests
$r_1,\ldots,r_n$ are drawn independently from a distribution $D$ on $X$, which is
not given to the algorithm. The algorithm observes $r_t$ before assigning it,
can retain all previously observed requests, and receives no samples from $D$
before time~$1$.

For a request sequence $R=(r_1,\ldots,r_n)$, define
\begin{equation}\label{eq:opt-def}
  \OPT(R,S)=\min_{\sigma\in\mathfrak S_n}\sum_{t=1}^n d(r_t,s_{\sigma(t)}),
\end{equation}
where $\mathfrak S_n$ is the set of permutations of $[n]$. We abbreviate this
random variable by $\OPT$ when the placement and request distribution are
understood. For an online algorithm, let $\ALG$ denote its total matching
cost. An algorithm is $c(n)$-competitive in the unknown-i.i.d.\ model if,
for every $S$ and $D$, its total cost satisfies
$\E[\ALG]\le c(n)\E[\OPT]$. This is a ratio-of-expectations guarantee, not a
bound on $\E[\ALG/\OPT]$ or on each realized request sequence.

\paragraph{Reduction to finite metric spaces.}
We prove the result for finite metric spaces. For an arbitrary metric space,
use the reduction in \cite{MNP06}: move each
request to a nearest server location, run the algorithm on the resulting
requests, and assign the original request to the same server as the transformed
one. The nearest server is chosen among all $n$ servers, whether used or
unused, with ties broken by a fixed order of labels.
% The transformed requests are still i.i.d.\ and lie in the finite set of server locations.
This converts a $\beta$-competitive guarantee into a $(2\beta+1)$-competitive guarantee, without changing its asymptotic order.
Therefore, we assume that $X$ is finite in \cref{sec:fair-bias,sec:comparison,sec:algorithm,sec:analysis}.

\subsection{Transport costs}\label{subsec:transport}
Let $\delta_x$ denote the point mass at $x$. For probability distributions
$\mu,\eta$ on the finite metric space $X$, let
\[
  \mu=\sum_{i\in I}p_i\delta_{x_i},
  \qquad
  \eta=\sum_{j\in J}q_j\delta_{y_j},
\]
where $I,J$ are finite index sets, $(p_i)_{i\in I}$ and $(q_j)_{j\in J}$
are probability vectors, and $x_i,y_j\in X$. The positions in either
representation need not be distinct. Define
\begin{equation}\label{eq:transport}
\begin{aligned}
  W_1(\mu,\eta)=\min_{\gamma\ge0}\quad
    &\sum_{i\in I}\sum_{j\in J}\gamma(i,j)d(x_i,y_j)\\
  \text{subject to}\quad
    &\sum_{j\in J}\gamma(i,j)=p_i &&(i\in I),\\
    &\sum_{i\in I}\gamma(i,j)=q_j &&(j\in J).
\end{aligned}
\end{equation}
The variable $\gamma(i,j)$ is the mass sent from source index $i$ at $x_i$
to destination index $j$ at $y_j$. A feasible matrix
$\gamma=(\gamma(i,j))_{i\in I,j\in J}$ is called a \emph{transport plan}.
Its row indexed by $i$ sums to $p_i$, and its column indexed by $j$ sums to
$q_j$.
When $I=J$ and $x_i=y_i$ for every $i$, a \emph{diagonal entry}
$\gamma(i,i)$ represents mass left at the same position.

Let $\nu$ be the uniform distribution over the $n$ server labels, and let $\widehat D_t$ be the empirical distribution of the first $t$ requests:
\begin{equation}\label{eq:nu-Dhat}
  \nu=\frac1n\sum_{j=1}^n\delta_{s_j},
  \qquad
  \widehat D_t=\frac1t\sum_{i=1}^t\delta_{r_i}.
\end{equation}

We will use the fact that common mass can be left at the same index in an
optimal plan; see \cref{lem:common-mass}.

The following lemma expresses the optimal matching cost as a transport cost.
\begin{lemma}\label{lem:matching-transport}
For every realized request sequence $R$,
\[
  nW_1(\widehat D_n,\nu)=\OPT(R,S).
\]
\end{lemma}
\begin{proof}
Sending mass $1/n$ along each edge of a perfect matching gives a feasible
transport plan, so $nW_1(\widehat D_n,\nu)\le\OPT(R,S)$.
For the reverse inequality, let $\pi$ be an optimal plan in
\eqref{eq:transport} from $\widehat D_n$ to $\nu$, with rows indexed by
requests $i\in[n]$ and columns by servers $j\in[n]$. Every row and column
sums to $1/n$, so $n\pi$ is doubly stochastic. A convex decomposition of
$n\pi$ into permutation matrices therefore contains a permutation of cost
at most $nW_1(\widehat D_n,\nu)$. This permutation gives a perfect matching,
proving the reverse inequality.
\end{proof}

For the analysis, we introduce two expected transport costs.
Let $I_1,\ldots,I_n$ be drawn independently from the uniform
distribution on $[n]$ and put $z_t=s_{I_t}$. This sequence of requests from $\nu$
is separate from the unknown request sequence. Define
\begin{equation}\label{eq:f-def}
  \widehat\nu_t=\frac1t\sum_{i=1}^t\delta_{z_i},
  \qquad
  f_t=\E[W_1(\widehat\nu_t,\nu)]
  \quad(1\le t\le n).
\end{equation}
For nonempty $A\subseteq[n]$, write
\[
  \nu_A=\frac1{|A|}\sum_{i\in A}\delta_{s_i}.
\]
Let $\cU_k$ be the uniform distribution over all $k$-element subsets of $[n]$,
and define
\begin{equation}\label{eq:g-def}
  g_k=\E_{A\sim\cU_k}[W_1(\nu_A,\nu)]
  \quad(1\le k\le n).
\end{equation}
In particular, $g_n=0$, although $f_n$ need not be zero. Both definitions
continue to distinguish co-located labels. If $\OPT_\nu$ denotes the offline
cost for $n$ i.i.d.\ requests from $\nu$, \cref{lem:matching-transport} gives
\begin{equation}\label{eq:uniform-opt}
  \E[\OPT_\nu]=nf_n.
\end{equation}

\section{The Fair-Bias algorithm}\label{sec:fair-bias}
We recall \FB\ \cite[Algorithm~1]{GGPW19} for requests drawn independently
from $\nu$. All numbered statements cited from
\cite{GGPW19} refer to its full version, arXiv:1904.09284v1.

\subsection{Definition of \FB}
Suppose $F\subseteq[n]$ is the current set of unused servers and $k=|F|$.
Fix an optimal plan $\rho_F$ from $\nu$ to $\nu_F$ by solving
\begin{equation}\label{eq:rho}
\begin{aligned}
  \min_{\rho_F\ge0}\quad
    &\sum_{j=1}^n\sum_{i\in F}\rho_F(j,i)d(s_j,s_i)\\
  \text{subject to}\quad
    &\sum_{i\in F}\rho_F(j,i)=\frac1n &&(j\in[n]),\\
    &\sum_{j=1}^n\rho_F(j,i)=\frac1k &&(i\in F).
\end{aligned}
\end{equation}
Rows index input server labels $j\in[n]$, and columns index available server
labels $i\in F$. The rule selecting an optimal plan depends only on $F$ and
$S$, not on the current input label. On input label $j$, \FB\ chooses server
$i\in F$ with probability $n\rho_F(j,i)$ and removes $i$ from $F$.
The optimal value in \eqref{eq:rho} is
$W_1(\nu,\nu_F)=W_1(\nu_F,\nu)$, denoted by $M(F)$ in \cite{GGPW19}.

\subsection{Properties and competitive-ratio bounds}
Run \FB\ on the labels $I_1,\ldots,I_n$ drawn independently from the uniform
distribution on $[n]$. Let $A_t$ be the
set of server labels it uses for the first $t$ requests, and let $L_t$ be the
total cost of those $t$ assignments.

\begin{lemma}[{A version of \cite[Lemma~2.2]{GGPW19}}]\label{lem:fb-prefix}
For every $t\in[n]$,
\begin{equation}\label{eq:prefix}
  A_t\sim\cU_t,
  \qquad
  \E[L_t]=\sum_{k=n-t+1}^n g_k.
\end{equation}
In particular, $\E[L_n]=\sum_{k=1}^n g_k$.
\end{lemma}
\begin{proof}
Condition on the past input labels and past internal random choices. The free
set $F$ is then fixed, and the next input label remains uniformly distributed
on $[n]$ and independent of that past. For each $i\in F$, its probability of being chosen next is
\[
  \sum_{j=1}^n\frac1n\,n\rho_F(j,i)=\frac1{|F|}.
\]
Hence the successive server labels are a uniformly random permutation of
$[n]$, and $A_t\sim\cU_t$.
Under the same conditioning, the expected next assignment cost is
\[
  \sum_{j=1}^n\frac1n\sum_{i\in F}n\rho_F(j,i)d(s_j,s_i)
  =W_1(\nu_F,\nu).
\]
When $|F|=k$, the unconditional distribution of $F$ is $\cU_k$, so the
unconditional expected cost of that step is $g_k$. The first $t$ steps use free
sets of sizes $n,n-1,\ldots,n-t+1$, giving the stated sum.
\end{proof}
The assertion $A_t\sim\cU_t$ is a statement about the distribution without
conditioning on the requests. It does not state that $A_t$ is independent of
$\widehat\nu_t$. In general these variables are dependent.

We also use the following competitive-ratio bounds.
\begin{proposition}[{\cite[Theorems~3.10 and~4.6]{GGPW19}}]\label{prop:fb-bound}
For every server placement, \FB\ with requests drawn independently
from $\nu$ satisfies
\begin{equation}\label{eq:fb-bound}
  \sum_{k=1}^n g_k=\E[L_n]\le\alpha(n)\E[\OPT_\nu],
  \qquad \alpha(n)=O((\log\log\log n)^2).
\end{equation}
For tree metrics, one may take $\alpha(n)=4$.
\end{proposition}

\section{Comparing cumulative transport costs}\label{sec:comparison}
We prove \eqref{eq:sum-comparison} using the execution of \FB\ on independent
requests from $\nu$ in \cref{sec:fair-bias}. Only \cref{lem:fb-prefix} is used in this section; the
competitive-ratio bound \eqref{eq:fb-bound} is not needed until the comparison
has been established.

We first relate $g_k$ to $g_{n-k}$.
The following identity is also used in the proof of
\cite[Lemma~3.9, Appendix~C.3]{GGPW19}.

\begin{lemma}\label{lem:complement}
If $1\le k<n$ and $A\subseteq[n]$ has size $k$, then, writing $A^c=[n]\setminus A$,
\begin{equation}\label{eq:complement}
  kW_1(\nu_A,\nu)=(n-k)W_1(\nu_{A^c},\nu).
\end{equation}
Consequently,
\begin{equation}\label{eq:g-complement}
  kg_k=(n-k)g_{n-k}.
\end{equation}
\end{lemma}
\begin{proof}
Use server labels on both sides of the transport plans.
By \cref{lem:common-mass}, an optimal plan from $\nu_A$ to $\nu$ may leave
mass $1/n$ at each label of $A$. After removing these zero-cost diagonal
entries, the remaining plan has entries $\gamma_{ij}$ only for $i\in A$,
$j\in A^c$, with
\begin{equation}\label{eq:complement-forward}
  \sum_{j\in A^c}\gamma_{ij}=\frac1k-\frac1n=\frac{n-k}{kn}
  \quad(i\in A),
  \qquad
  \sum_{i\in A}\gamma_{ij}=\frac1n
  \quad(j\in A^c).
\end{equation}
Conversely, any nonnegative matrix $(\gamma_{ij})_{i\in A,j\in A^c}$
satisfying \eqref{eq:complement-forward} can be completed to a feasible plan
by restoring those diagonal entries. Thus $W_1(\nu_A,\nu)$ is exactly the
minimum cost over such matrices.

For transport from $\nu_{A^c}$ to $\nu$, leave mass $1/n$ at
each label of $A^c$. After removing these diagonal entries, the remaining
entries $\eta_{ji}$ must satisfy
\begin{equation}\label{eq:complement-backward}
  \sum_{i\in A}\eta_{ji}=\frac1{n-k}-\frac1n=\frac{k}{n(n-k)}
  \quad(j\in A^c),
  \qquad
  \sum_{j\in A^c}\eta_{ji}=\frac1n
  \quad(i\in A).
\end{equation}
Given a feasible $\gamma$ in \eqref{eq:complement-forward}, set
\[
  \eta_{ji}=\frac{k}{n-k}\gamma_{ij}.
\]
Its row and column sums are, respectively,
\[
  \sum_{i\in A}\eta_{ji}
    =\frac{k}{n-k}\cdot\frac1n=\frac{k}{n(n-k)},
  \qquad
  \sum_{j\in A^c}\eta_{ji}
    =\frac{k}{n-k}\cdot\frac{n-k}{kn}=\frac1n.
\]
It is therefore feasible for \eqref{eq:complement-backward}. By symmetry of
the distance, its cost is
\[
  \sum_{j\in A^c}\sum_{i\in A}\eta_{ji}d(s_j,s_i)
   =\frac{k}{n-k}\sum_{i\in A}\sum_{j\in A^c}\gamma_{ij}d(s_i,s_j).
\]
The inverse transformation is $\gamma_{ij}=((n-k)/k)\eta_{ji}$. Its feasibility
follows by checking the row and column sums in the same way. Hence the two
optimal residual costs differ by exactly the factor $k/(n-k)$, proving
\eqref{eq:complement}. Finally, complementation maps a uniform $k$-element
subset to a uniform $(n-k)$-element subset. Taking expectations gives
\eqref{eq:g-complement}.
\end{proof}

We next bound $f_t$ using the first $t$ assignments made by \FB.
\begin{lemma}\label{lem:per-time}
For every $t\in[n]$,
\begin{equation}\label{eq:per-time}
  f_t\le g_t+\frac1t\sum_{k=n-t+1}^n g_k.
\end{equation}
\end{lemma}
\begin{proof}
Consider the execution of \FB\ on independent requests from $\nu$ defining
$A_t$ and $L_t$ in \cref{lem:fb-prefix}. For each realization, send mass $1/t$ along each of its
first $t$ assignment edges. The source distribution is $\widehat\nu_t$;
the destination distribution is $\nu_{A_t}$, because these $t$ servers are
distinct. Thus
\[
  W_1(\widehat\nu_t,\nu_{A_t})\le\frac{L_t}{t},
\]
and the triangle inequality gives
\[
  W_1(\widehat\nu_t,\nu)
    \le\frac{L_t}{t}+W_1(\nu_{A_t},\nu).
\]
Take expectation over this same execution, including the requests drawn from
$\nu$ and the internal random choices. The left-hand expectation is $f_t$, since that
term depends only on the requests. The last term depends only on $A_t$, so its
expectation is $g_t$ by the unconditional distribution $A_t\sim\cU_t$.
Using \cref{lem:fb-prefix} for the middle term yields
\[
  f_t\le\frac1t\E[L_t]+g_t
     =\frac1t\sum_{k=n-t+1}^n g_k+g_t.
\]
Neither taking expectations nor evaluating these individual terms requires
independence between $A_t$ and $\widehat\nu_t$.
\end{proof}

Summing the preceding bounds over time and using \cref{lem:complement}
gives the following theorem.
\begin{theorem}\label{thm:sum-comparison}
For every server placement,
\begin{equation}\label{eq:sum-comparison}
  \sum_{t=1}^n f_t\le 2\sum_{k=1}^n g_k.
\end{equation}
\end{theorem}
\begin{proof}
The claim is immediate for $n=1$, since $f_1=g_1=0$. For $n\ge2$, sum
\eqref{eq:per-time} over $t$:
\begin{equation}\label{eq:sum-first}
  \sum_{t=1}^n f_t
  \le\sum_{t=1}^n g_t
     +\sum_{t=1}^n\frac1t\sum_{k=n-t+1}^n g_k.
\end{equation}
Since $g_n=0$, reversing the order of summation gives
\[
  \sum_{t=1}^n\frac1t\sum_{k=n-t+1}^n g_k
    =\sum_{k=1}^{n-1}g_k\sum_{t=n-k+1}^n\frac1t.
\]
The inner sum has $k$ terms, each at most $1/(n-k+1)$. By
\eqref{eq:g-complement},
\begin{align*}
  \sum_{k=1}^{n-1}g_k\sum_{t=n-k+1}^n\frac1t
  &\le\sum_{k=1}^{n-1}\frac{k}{n-k+1}g_k\\
  &=\sum_{k=1}^{n-1}\frac{n-k}{n-k+1}g_{n-k}\\
  &=\sum_{j=1}^{n-1}\frac{j}{j+1}g_j
   \le\sum_{j=1}^n g_j.
\end{align*}
Substituting into \eqref{eq:sum-first} proves the theorem.
\end{proof}
Combining the theorem with \eqref{eq:fb-bound} and \eqref{eq:uniform-opt} gives
\begin{equation}\label{eq:f-alpha}
  \sum_{t=1}^n f_t\le2\alpha(n)\E[\OPT_\nu]
    =2\alpha(n)nf_n.
\end{equation}

\section{Generating virtual requests}\label{sec:algorithm}
We return to the actual requests $r_1,\ldots,r_n$ from the unknown distribution
$D$. The algorithm uses two different transport plans: an optimal transport
plan $\pi_t$ from $\widehat D_t$ to $\nu$ that generates a virtual request
$r'_t$ from the actual request $r_t$, and the plan $\rho_F$ used by \FB\ to
match that virtual request to an available server.

\subsection{The algorithm and plan-selection rules}\label{subsec:virtual-generation}
At time $t$, choose an optimal plan $\pi_t$ from $\widehat D_t$ to $\nu$
in \eqref{eq:transport}, with rows indexed by requests $i\in[t]$ at $r_i$
and columns by server labels $j\in[n]$ at $s_j$. Thus
\[
  \sum_{j=1}^n\pi_t(i,j)=\frac1t\quad(i\in[t]),
  \qquad
  \sum_{i=1}^t\pi_t(i,j)=\frac1n\quad(j\in[n]).
\]
We use any fixed tie-breaking rule that selects the same transport plan
whenever the request locations coincide as multisets.

Having observed $r_t$, draw a virtual label $J_t\in[n]$ with probabilities
\begin{equation}\label{eq:virtual-law}
  \Prob(J_t=j\mid r_1,\ldots,r_t)
   =t\pi_t(t,j),
  \qquad j\in[n],
\end{equation}
and set $r'_t=s_{J_t}$. These probabilities are nonnegative and sum to one
because row $t$ of $\pi_t$ sums to $1/t$.
The draw uses fresh randomness independent of the request sequence
and the randomness used to draw the virtual labels at other times.

Pass $J_t$ to \FB, whose plan $\rho_F$ is allowed to depend on the current
free set $F$. Use randomness for \FB\ that is independent of the request
sequence and of all randomness used to draw the virtual labels. If \FB\ selects
label $a_t$, assign the actual request $r_t$ to that same server. \FB\ bases
its decisions only on its virtual inputs, its own state, and its own randomness;
it does not inspect the original request history.

\begin{algorithm}[t]
\caption{Online metric matching for unknown i.i.d.\ requests in a finite metric space}\label{alg:main}
\begin{algorithmic}[1]
\Require Server placement $S=(s_1,\ldots,s_n)$ and the fixed plan-selection rules
\State $F\gets[n]$
\For{$t=1,\ldots,n$}
  \State Observe $r_t$ and form $\widehat D_t=t^{-1}\sum_{\ell=1}^t\delta_{r_\ell}$.
  \State Choose the fixed optimal plan $\pi_t$ from $\widehat D_t$ to $\nu$
         in \eqref{eq:transport}, using request indices $i\in[t]$ and server
         labels $j\in[n]$.
  \State Draw $J_t=j$ with probability $t\pi_t(t,j)$.
  \State Choose the fixed \FB\ plan $\rho_F$ in \eqref{eq:rho}.
  \State Draw $a_t=i\in F$ with probability $n\rho_F(J_t,i)$.
  \State Assign $r_t$ irrevocably to server $a_t$ and set $F\gets F\setminus\{a_t\}$.
\EndFor
\end{algorithmic}
\end{algorithm}
The virtual label $J_t$ does not specify the final assignment and may refer to
a server that has already been used. The transport plans can be computed
using only the observed request positions and the known server locations.

\subsection{Independence of the virtual sequence}
We first show that the virtual requests $r'_1,\ldots,r'_n$ are independent samples from $\nu$.
For a fixed multiset of requests $M=\{x_1,\ldots,x_t\}$, write $\pi_M(i,j)$
for the selected plan, with row $i$ corresponding to request $x_i$.

\begin{lemma}\label{lem:uniform-virtual}
Fix $t\le n$ and any multiset $M$ of $t$ requests. Present these requests
in a uniformly random order and apply the conversion rule above.
For every $j_1,\ldots,j_t\in[n]$,
\begin{equation}\label{eq:virtual-joint}
  \Prob(J_1=j_1,\ldots,J_t=j_t)=n^{-t}.
\end{equation}
Consequently, for i.i.d.\ actual requests, $J_1,\ldots,J_n$ are independent
and each is uniformly distributed on $[n]$, and $r'_1,\ldots,r'_n$ are independent
samples from $\nu$.
\end{lemma}
\begin{proof}
We prove the lemma by induction on $t$, simultaneously over all multisets of size $t$. For $t=1$,
the plan has $\pi_M(1,j)=1/n$ for every server label $j$, so the base case
follows.

For the inductive step, fix $M=\{x_1,\ldots,x_t\}$. Let $K$ be the index of
the final request, so $\Prob(K=i)=1/t$ for every $i\in[t]$. Conditional
on $K=i$, the preceding requests are a uniformly random ordering of the
remaining $t-1$ requests. By the induction hypothesis,
\begin{equation}\label{eq:induction-prefix}
  \Prob(J_1=j_1,\ldots,J_{t-1}=j_{t-1}\mid K=i)
    =n^{-(t-1)}.
\end{equation}
By the conversion rule, we have
\begin{equation}\label{eq:induction-last}
  \Prob(J_t=j_t\mid K=i,J_1=j_1,\ldots,J_{t-1}=j_{t-1})
    =t\pi_M(i,j_t).
\end{equation}
Therefore,
\begin{align*}
  \Prob(J_1=j_1,\ldots,J_t=j_t)
    &=\sum_{i=1}^t\frac1t\,n^{-(t-1)}\,t\pi_M(i,j_t)
    =n^{-(t-1)}\sum_{i=1}^t\pi_M(i,j_t)=n^{-t}.
\end{align*}
The last equality is the destination constraint for server label $j_t$.
This completes the induction.
\end{proof}
By \cref{lem:uniform-virtual}, the virtual requests $r'_1,\ldots,r'_n$
generated in \cref{subsec:virtual-generation} are i.i.d.\ samples from $\nu$.
They are also independent of the internal randomness of \FB, so the analysis
in \cref{sec:fair-bias} applies to the execution of \FB\ on the virtual requests.

\subsection{The conversion cost}
We next relate the expected cost of converting an actual request to a virtual
request to the expected transport cost between $\widehat D_t$ and $\nu$.

\begin{lemma}\label{lem:conversion}
For each $t\in[n]$,
\begin{equation}\label{eq:conversion-cost}
  \E[d(r_t,r'_t)]=\E[W_1(\widehat D_t,\nu)].
\end{equation}
\end{lemma}
\begin{proof}
Let $M_t=\{r_1,\ldots,r_t\}$ be the multiset of the first $t$ actual requests.
Conditioned on $M_t=M=\{x_1,\ldots,x_t\}$, each request has probability $1/t$
of being the final request. When request $x_i$ is last, its virtual label
is drawn with probabilities $t\pi_M(i,j)$. Thus
\begin{align*}
  \E[d(r_t,r'_t)\mid M_t=M]
  &=\frac1t\sum_{i=1}^t\sum_{j=1}^n t\pi_M(i,j)d(x_i,s_j)\\
  &=\sum_{i=1}^t\sum_{j=1}^n\pi_M(i,j)d(x_i,s_j)
    =W_1(\widehat D_t,\nu).
\end{align*}
The last equality uses optimality of $\pi_M$. Averaging over $M_t$ yields the lemma.
\end{proof}
Let $\ALG_\nu=\sum_{t=1}^n d(r'_t,s_{a_t})$ be the cost of running \FB\ on the
virtual requests. For each realization, the triangle inequality yields
$d(r_t,s_{a_t})\le d(r_t,r'_t)+d(r'_t,s_{a_t})$.
Summing and using \cref{lem:conversion,lem:uniform-virtual,lem:fb-prefix}, we obtain
\begin{equation}\label{eq:alg-preliminary}
  \E[\ALG]
  \le\sum_{t=1}^n\E[W_1(\widehat D_t,\nu)]+\E[\ALG_\nu]
  =\sum_{t=1}^n\E[W_1(\widehat D_t,\nu)]+\sum_{k=1}^n g_k.
\end{equation}

\section{Competitive analysis}\label{sec:analysis}
We now compare the two terms in \eqref{eq:alg-preliminary} with the expected
offline optimum for the actual requests. The unknown distribution $D$ is used
only in the analysis.

\begin{lemma}\label{lem:distribution-comparison}
For every server placement and request distribution $D$,
\begin{align}
  nW_1(D,\nu)&\le\E[\OPT],\label{eq:distribution-lb}\\
  \E[W_1(\widehat D_t,\nu)]&\le W_1(D,\nu)+f_t
      &&(1\le t\le n),\label{eq:empirical-comparison}\\
  \E[\OPT_\nu]&\le\E[\OPT]+nW_1(D,\nu)
                \le2\E[\OPT].\label{eq:opt-comparison}
\end{align}
\end{lemma}
\begin{proof}
Write $X=\{x_1,\ldots,x_m\}$ and
$D=\sum_{a=1}^m p_a\delta_{x_a}$.
For \eqref{eq:distribution-lb}, fix an optimal matching $\sigma_R$ for each
realization $R$ and define a plan on $[m]\times[n]$ by
\[
  \beta_R(a,j)=\frac1n\sum_{i=1}^n
    \ind\{r_i=x_a,\ \sigma_R(i)=j\}.
\]
The expected plan $\E[\beta_R]$ has row sums $p_a$ and column sums $1/n$,
so it is a feasible plan from $D$ to $\nu$. Since its cost is $\E[\OPT]/n$, \eqref{eq:distribution-lb} is proved.

Next fix an optimal plan $\gamma$ from $D$ to $\nu$ in \eqref{eq:transport},
with source indices $a\in[m]$ at $x_a$ and destination indices $j\in[n]$ at
$s_j$. Independently for each $i\in[n]$, draw an index pair
$(U_i,V_i)\in[m]\times[n]$ with probabilities
\[
  \Prob(U_i=a,V_i=j)=\gamma(a,j)
  \quad(a\in[m],\ j\in[n]),
\]
and set $\widetilde r_i=x_{U_i}$ and $z_i=s_{V_i}$.
The sequence $(\widetilde r_i)_{i \in [n]}$ is i.i.d.\ from $D$, and $(z_i)_{i \in [n]}$ is i.i.d.\ from $\nu$.
For each $t$, the empirical distribution of $\widetilde r_1,\ldots,\widetilde r_t$ follows $\widehat D_t$. In addition, the empirical distribution of $z_1,\ldots,z_t$ follows $\widehat\nu_t$ and we have $\E[d(\widetilde r_i,z_i)]=W_1(D,\nu)$.
For each realization and each $t$, define a plan $\lambda_t$ with source $\widetilde r_i$ and destination $z_j$ by
\begin{equation}\label{eq:paired-plan}
  \lambda_t(i,j)=\frac1t\ind\{i=j\}.
\end{equation}
Every row and column sums to $1/t$, so this is a feasible plan from
$\widehat D_t$ to $\widehat\nu_t$. Its cost is
$t^{-1}\sum_{i=1}^t d(\widetilde r_i,z_i)$, so
\begin{equation}\label{eq:paired-bound}
  W_1(\widehat D_t,\widehat\nu_t)
   \le\frac1t\sum_{i=1}^t d(\widetilde r_i,z_i).
\end{equation}
Adding transport from $\widehat\nu_t$ to $\nu$ and taking expectations gives
\[
  \E[W_1(\widehat D_t,\nu)]\le W_1(D,\nu)+f_t,
\]
proving \eqref{eq:empirical-comparison}.

Finally, let $\sigma$ be an optimal matching of
$(\widetilde r_1,\ldots,\widetilde r_n)$ to $S$. Match each $z_i$ to
$s_{\sigma(i)}$ as well. This is feasible, because the server labels
$\sigma(i)$ are all distinct. Therefore,
\begin{align*}
  \OPT((z_1,\ldots,z_n),S)
  &\le\sum_{i=1}^n d(z_i,s_{\sigma(i)})
  \le\sum_{i=1}^n d(z_i,\widetilde r_i)
     +\OPT((\widetilde r_1,\ldots,\widetilde r_n),S).
\end{align*}
Taking expectations and using \eqref{eq:distribution-lb} proves
\eqref{eq:opt-comparison}.
\end{proof}

We now combine these inequalities with \cref{thm:sum-comparison}
to obtain the competitive-ratio bound.
\begin{theorem}\label{thm:core-reduction}
Suppose \FB\ is $\alpha(n)$-competitive for requests drawn independently from
$\nu$ on the server placements under consideration. Then \cref{alg:main} satisfies
\begin{equation}\label{eq:main-reduction}
  \E[\ALG]\le(1+6\alpha(n))\E[\OPT]
\end{equation}
for every unknown i.i.d.\ request distribution on a finite metric space.
\end{theorem}
\begin{proof}
By \eqref{eq:alg-preliminary}, \eqref{eq:empirical-comparison}, and
\cref{thm:sum-comparison},
\begin{align*}
  \E[\ALG]
  &\le\sum_{t=1}^n\E[W_1(\widehat D_t,\nu)]+\sum_{k=1}^n g_k
  \le nW_1(D,\nu)+\sum_{t=1}^n f_t+\sum_{k=1}^n g_k\\
  &\le nW_1(D,\nu)+3\sum_{k=1}^n g_k
  \le nW_1(D,\nu)+3\alpha(n)\E[\OPT_\nu]\\
  &\le\E[\OPT]+6\alpha(n)\E[\OPT].
\end{align*}
The last line uses \eqref{eq:distribution-lb} and \eqref{eq:opt-comparison}.
\end{proof}

\begin{proof}[Proof of \cref{thm:main}]
Combining \cref{thm:core-reduction,prop:fb-bound} gives the stated asymptotic
ratio on finite metric spaces. The reduction in \cref{subsec:model} gives
the same asymptotic ratio on arbitrary metric spaces.
\end{proof}

For tree metrics, substituting $\alpha(n)=4$ from
\cref{prop:fb-bound} into \cref{thm:core-reduction}
gives a constant competitive ratio.
\begin{corollary}\label{cor:trees}
On tree metrics, unknown i.i.d.\ requests can be served with competitive
ratio $51$, with no restriction on the support of the request distribution.
If the request distribution has finite support, \cref{alg:main} is
$25$-competitive.
\end{corollary}
\begin{proof}
If $D$ has finite support, its support together with the server locations
forms a finite metric space, so \cref{thm:core-reduction} gives $1+6\cdot4=25$.
The reduction in \cref{subsec:model} gives $2\cdot25+1=51$ for arbitrary
request distributions.
\end{proof}

\section{Comparison with random-order arrivals}\label{sec:consequences}
As summarized in \cref{tab:models}, the upper bounds in
\cref{thm:main,cor:trees} are asymptotically smaller than the random-order
lower bound \cite{Rag16}.

\Cref{lem:uniform-virtual} holds for a uniformly random ordering of any fixed
multiset. It is therefore important to identify why the competitive analysis
cannot simply be repeated for random-order inputs. The obstruction is the
comparison in \eqref{eq:opt-comparison} between the offline optimum for $n$
independent requests from $\nu$ and the offline optimum for the actual requests:
exchangeability alone does not imply that comparison.

For example, take two servers at $0$ and $1$ with distance one, and let the
random-order request multiset be $M=\{0,1\}$. Its empirical distribution is
$D_M=\nu$, and
\[
  \OPT(M,S)=0,\qquad W_1(D_M,\nu)=0.
\]
However, two independent requests from $\nu$ coincide with probability $1/2$;
when they do, their optimal matching cost is one. Hence
\[
  \E[\OPT_\nu]=\frac12
   >\OPT(M,S)+nW_1(D_M,\nu)=0.
\]
Thus the counterpart of \eqref{eq:opt-comparison} is false in the random-order
model. The exact generation of independent virtual inputs by itself is not a
competitive-ratio guarantee; the conversion cost must also be charged to an
appropriate offline benchmark.

\section*{AI Disclaimer}
OpenAI's Codex and ChatGPT 6 Astra were used to assist with revising the text
and notation and discussing the proofs. The author is responsible for
the final content of this manuscript.

\bibliographystyle{alphaurl}
\bibliography{references}

\appendix
\crefalias{section}{appendix}
\section{Leaving common mass in place}\label{app:transport}
We use the matrix representation of transport plans from \cref{subsec:transport}.

\begin{lemma}\label{lem:common-mass}
Let $(p_i)$ and $(q_i)$ be nonnegative masses of the same total mass on a
finite label set. With transport cost $d(s_i,s_j)$, an optimal plan may be
chosen so that
\[
  \gamma(i,i)=\min\{p_i,q_i\}\quad\text{for every }i.
\]
\end{lemma}
\begin{proof}
The feasible set of transport plans is nonempty and compact. Among its
cost-minimizing plans, choose one maximizing the total diagonal mass
$\sum_i\gamma(i,i)$. If at some label $v$,
$\gamma(v,v)<\min\{p_v,q_v\}$, then there exist $u\ne v$ and $w\ne v$ with
$\gamma(u,v)>0$ and $\gamma(v,w)>0$. Let
\[
  \varepsilon=\min\{\gamma(u,v),\gamma(v,w)\}>0.
\]
Decrease $\gamma(u,v)$ and $\gamma(v,w)$ by $\varepsilon$, and increase
$\gamma(u,w)$ and $\gamma(v,v)$ by $\varepsilon$. The source row sums and
destination column sums are unchanged, and no entry becomes negative. The
cost change is
\[
  \varepsilon\bigl(d(s_u,s_w)-d(s_u,s_v)-d(s_v,s_w)\bigr)\le0.
\]
Thus the modified plan is still optimal. Its diagonal mass has increased by
$\varepsilon$, or by $2\varepsilon$ if $u=w$, contradicting the choice of
$\gamma$. Hence all diagonal entries equal the common mass.
\end{proof}
After that common mass is removed, only labels with excess source mass need
send mass, and only labels with excess destination mass need receive it.
This is the residual transport used in \cref{lem:complement}.

\end{document}